\documentclass[10pt,prl,aps,twocolumn,showpacs,floatfix]{revtex4}

\usepackage{graphicx}
\usepackage{amssymb}
\usepackage{subfigure}
\usepackage{color}
\usepackage{amsmath} 
\newcommand{\be}{\begin{eqnarray}}
\newcommand{\ee}{\end{eqnarray}}

\begin{document}

\title{Ashkin-Teller criticality and pseudo first-order behavior \\ 
in a frustrated Ising model on the square lattice}

\author{Songbo Jin, Arnab Sen, and Anders W. Sandvik}
\affiliation{Department of Physics, Boston University, 590 Commonwealth Avenue, Boston, Massachusetts 02215, USA}

\begin{abstract}
We study the challenging thermal phase transition to stripe order in the frustrated square-lattice Ising model with couplings $J_1<0$ (nearest-neighbor, 
ferromagnetic) and $J_2>0$ (second-neighbor, antiferromagnetic) for $g=J_2/|J_1|>1/2$. Using Monte Carlo simulations and known analytical results, we 
demonstrate Ashkin-Teller criticality for $g \ge g^*$, i.e., the critical exponents vary continuously between those of the $4$-state Potts model at $g=g^*$ 
and the Ising model for $g \to \infty$. Thus, stripe transitions offer a route to realizing a related class of conformal field theories with conformal 
charge $c=1$ and varying exponents. The transition is first-order for $g<g^*= 0.67 \pm 0.01$, much lower than previously believed, and exhibits 
{\it pseudo first-order} behavior for $g^* \le g \alt 1$.
\end{abstract}

\date{\today}


\maketitle

Breaking a four-fold symmetry (Z$_4$) at a thermal phase transition in a two-dimensional (2D) system can lead to a very special kind of critical behavior, where 
the critical exponents depend on microscopic details, in contrast to the universal exponents at normal transitions. Continuously varying exponents are possible in 
conformal field theory (CFT) when the conformal charge $c\ge 1$ \cite{Friedan,Cardybook}. A well-known microscopic Z$_4$-symmetric model realizes this behavior for
$c=1$; the Ashkin-Teller (AT) model \cite{AToriginal,AT} consisting of two square-lattice Ising models coupled by a four-spin interaction. A Z$_4$-symmetric order 
parameter also obtains in the standard ferromagnetic Ising model when sufficiently strong frustrating (antiferromagnetic) second-neighbor couplings are added, in which 
case the low-temperature ordered state exhibits stripes of alternating magnetization (which can be arranged in four different ways on the square lattice). While $c=1$ 
AT criticality has been suspected at the stripe transition, it has not been possible to demonstrate this convincingly for a wide range of couplings (only at very 
strong second-neighbor coupling, where the system reduces to two weakly coupled Ising models \cite{honecker1,honecker2}). It has even been difficult to determine whether 
the transition is continuous or first-order \cite{Nightingale,Swendsen,Oitmaa,Binder,Landau,LandauBinder,variational1,variational2,honecker1,honecker2}. A convincing 
demonstration of robust $c=1$ criticality with varying exponents would be important, since stripe ordering can be achieved in many different systems \cite{Edlund}
and represents a more realistic route to $c=1$ and varying exponents than the somewhat artificial AT model. 

Using Monte Carlo (MC) simulations of the frustrated Ising and AT models, along with known analytical results for the AT model, we here 
demonstrate $c=1$ scaling at the stripe transition, with continuously varying exponents matching those of the AT model. We also uncover a 
remarkable {\it pseudo first-order behavior} associated with this type of continuous phase transition, which had not been previously noticed and which led to wrong 
conclusions regarding the nature of the stripe transition.

The 2D frustrated Ising model with couplings $J_1<0$ (ferromagnetic) and $J_2>0$ (antiferromagnetic) is defined by the Hamiltonian
\be 
H = J_1\sum_{\langle ij \rangle}\sigma_i \sigma_j + J_2 \sum_{\langle \langle ij \rangle \rangle}\sigma_i \sigma_j,
\label{eq1}
\ee
where first and second (diagonal) neighbors on the square lattice are denoted by $\langle ij \rangle$ and $\langle \langle ij \rangle \rangle$, respectively, 
and $\sigma_i = \pm 1$. A stripe state obtains at low temperature ($T$) when the coupling ratio $g=J_2/|J_1| >1/2$. Even though this model represents one of the 
most natural and simple extensions of the standard 2D Ising model, its stripe transition remains highly controversial. The most basic question under debate is whether 
the transition is continuous for all $g>1/2$, or 
there is a line of first-order transitions up to a point $g^*$. The next issue is to establish the universality class of the continuous transitions. Early 
numerical and analytic approaches supported the idea that the transition is always continuous, but with the critical exponents changing with $g$. Some variational 
studies \cite{variational1,variational2} and recent MC studies \cite{honecker1,honecker2} have, however, found a line of first-order transitions for $1/2<g\lesssim 1$. 
A very recent MC study \cite{honecker2} used a double-peak structure in energy histograms to conclude that the transition is first-order at least up to $g=0.9$. 
For large $g$ the transition should be AT-like, based on an effective continuum theory of two weakly coupled Ising models \cite{honecker2}. For smaller $g$, 
the AT behavior could either continue down to the point at which the transition turns first-order (with the critical end-point of the AT line corresponding to 
the 4-state Potts model), or some other behavior that is beyond the AT description may apply instead. The MC results of Ref.~\onlinecite{honecker2} could not 
distinguish between these scenarios.

We show here that the stripe transition is first order in a much smaller range of couplings than claimed in Ref.~\onlinecite{honecker2}; for $1/2<g<g^*$ with 
$g^* \approx 0.67$. For $g>g^*$ it is continuous and in the AT class. The exponents change continuously with $g$ as in the AT model \cite{AT}, with $g^*$ 
corresponding to the $4$-state Potts model \cite{Baxter,Salas} and 
$g \rightarrow \infty$ to standard Ising universality \cite{AT}. First-order indicators (necessary but not sufficient), e.g., multiple peaks in energy 
and order-parameter distributions, lead to over-estimation of the region of discontinuous 
transitions. As we will show, this is because the $4$-state Potts model and neighboring transitions in the AT model exhibit pseudo first-order behavior, 
though these transitions are known to be continuous \cite{AT}. The Potts point had not been identified in previous works, and the pseudo first-order behavior 
was interpreted as actual first-order transitions.

{\it Stripe transition.}---The striped phase is characterized by a two-component order parameter $(m_x,m_y)$ with 
\be 
m_x = \frac{1}{N}\sum_{i=1}^N\sigma_i(-1)^{x_i},\mbox{~~~} m_y = \frac{1}{N}\sum_{i=1}^N\sigma_i(-1)^{y_i},
\label{eq2}
\ee
where $(x_i,y_i)$ are the coordinates of site $i$ on an $L \times L$ periodic lattice and $N=L^2$. We define $m^2=m_x^2+m_y^2$ and the stripe susceptibility 
$\chi = N(\langle m^2\rangle - \langle |m|\rangle^2)/T$. We employ the standard single-spin Metropolis MC algorithm and use $|J_1|=1$ as the unit of $T$.

\begin{figure}
\center{\includegraphics[width=8.2cm, clip]{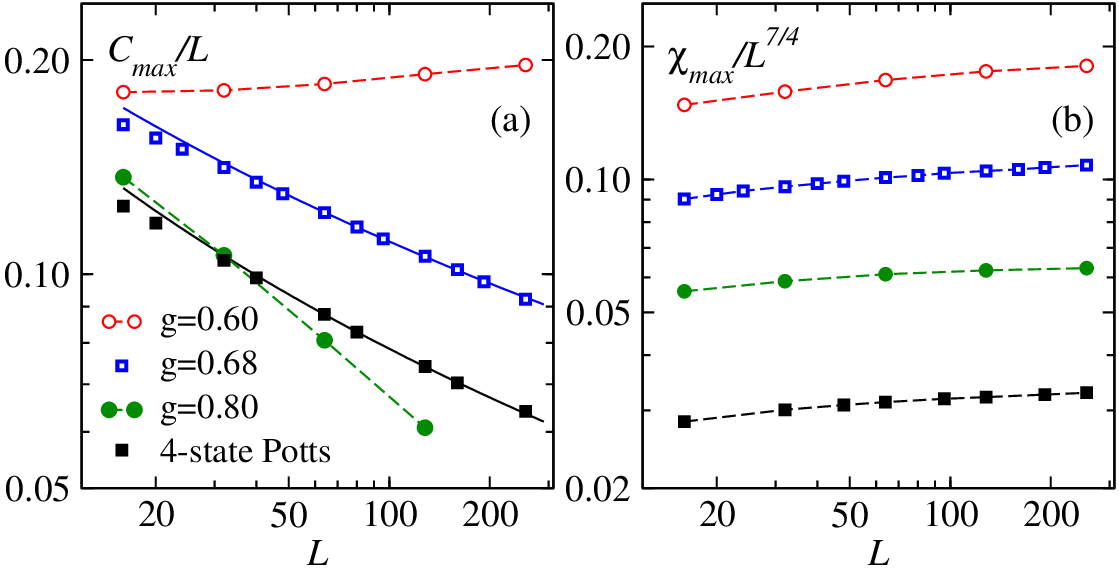}}
\vskip-1mm
\caption{(Color online) Peak value of the specific heat and stripe susceptibility vs $L$ for the $J_1$-$J_2$ and $4$-state Potts models. Factors
corresponding the asymptotic Potts scaling have been divided out. In (a) the curves through the $g=0.68$ and Potts data are fits to the Potts 
form $C=aL\ln(L/b)^{-3/2}$. A log-correction should also be present in (b) but the expected Potts form $\chi=aL^{7/4}\ln(L/b)^{-1/8}$ \cite{Salas}
is not seen for $L\le 256$.}
\label{fig1}
\vskip-3mm
\end{figure}

We analyze the peak value $C_{\rm max}(L)$ of the specific heat and $\chi_{\rm max}(L)$ of the stripe susceptibility. By standard finite-size scaling arguments \cite{Goldenfeld}, 
$C_{\rm max}(L) \sim L^{\alpha/\nu}$ and $\chi_{\rm max}(L) \sim L^{\gamma/\nu}$. For first-order 2D transitions these quantities should instead diverge as $L^2$. 
Examples of the scaling behavior are shown in Fig~\ref{fig1}. For the $L \le 256$ systems studied, the exponent $\alpha/\nu$, estimated from the slope 
on the log-log scale, decreases with increasing $g$ (it is close to $2$ for $g=0.51$), while $\gamma/\nu$ remains close to $7/4$ for the $g$-values 
displayed here (also approaching $2$ closer to $g=1/2$). These behaviors can be affected by scaling corrections, and the slopes for moderate sizes may 
not reflect the true exponents. Fig.~\ref{fig1} also shows results for the $4$-state Potts model, for which it is rigorously known that $\alpha/\nu=1$ and 
$\gamma/\nu=7/4$, but there are multiplicative logarithmic scaling corrections that affect the behavior strongly for lattices accessible in MC simulations
\cite{Salas}. For $g$ in the range $0.66 - 0.70$, the $J_1$-$J_2$ scaling agrees well (up to factors) with that of the Potts model.

According to CFT \cite{Friedan,Cardybook}, the exponents can change continuously when the ordered phase breaks Z$_4$ symmetry. There are two known microscopic 
scenarios, exemplified by: (i) The XY model in a four-fold anisotropy field $h_4$. For $h_4=0$ there is a Kosterlitz-Thouless transition \cite{Berezinskii,KT}, 
while $|h_4| \rightarrow \infty$ gives standard Ising universality \cite{JKKN}. (ii) The line of fixed points connecting the Ising and $4$-state Potts points in 
the square-lattice AT model \cite{AToriginal,AT}. Assuming that one of these scenarios applies to the continuous transitions in the $J_1$-$J_2$ model, 
$\gamma/\nu=7/4$ is fixed because this holds always for both (i) and (ii).
This explains why $\chi/L^{7/4}$ is almost constant for all cases in Fig.~\ref{fig1}(b). However, the observed variation of $\alpha/\nu$ with $g$ in 
Fig.~\ref{fig1}(a) (and at larger $g$) rules out scenario (i), since $\alpha=0$ for all the fixed points there. Then the natural scenario left to 
consider is that there is a line of fixed points corresponding the AT model. 

The AT Hamiltonian can be written as
\be
H = -\sum_{\langle ij \rangle} (\sigma_i \sigma_j + \tau_i \tau_j + K\sigma_i \sigma_j \tau_i \tau_j),
\label{eq5}
\ee
where two Ising variables, $\sigma_i,\tau_i$, reside on each site $i$ and are coupled to each other through $K$. The ferromagnetic phase of the AT model 
($\langle \sigma \tau \rangle \neq 0$ and $\langle \sigma \rangle = \pm \langle \tau \rangle$) breaks Z$_4$ symmetry and the order parameter can 
again be expressed as a vector $(m_{\sigma},m_{\tau})$ with
\be
m_{\sigma} = \frac{1}{N}\sum_{i=1}^N \sigma_i, \mbox{~~~~} m_{\tau}=\frac{1}{N}\sum_{i=1}^N \tau_i.
\label{eq6}
\ee
The relevant Z$_4$ transitions take place when $K \in [0,1]$, where $K=0$ corresponds to two decoupled Ising models and $K=1$ can be mapped to the $4$-state 
Potts model. The  $T>0$ transitions for $K \in [0,1]$ are all continuous, with the exponents depending on $K$ (but $\eta=1/4$ always) \cite{AT}. 

{\it Binder cumulant.}---To further understand the nature of the phase transitions, we probe the Binder cumulant of the stripe order parameter. For a
$2$-component vector order-parameter, the cumulant is defined as
\be 
U = 2 \left( 1-\frac{1}{2}\frac{\langle m^4 \rangle}{\langle m^2 \rangle^2}\right),
\label{eq4}
\ee
where the factors are chosen to make $U \rightarrow 0$ in the disordered phase and $U \rightarrow 1$ in the ordered phase in the thermodynamic 
limit. We define $U$ for the AT model in the same way, by using $m^2=m_{\sigma}^2+m_{\tau}^2$. 

For continuous transitions, the cumulant typically grows monotonically upon lowering $T$ and stays bounded within $[0,1]$. It approaches a step 
function at $T_c$ as $L \to \infty$ \cite{Binder_c}. For a first-order transition, it instead shows a non-monotonic behavior with $T$ for large 
systems \cite{Vollmayr}, developing a negative peak which approaches $T_c$ and grows narrower and diverges as $L^2$ (in two dimensions) when $L \to \infty$.
The nonmonotonic behavior can be traced to the emergence of multiple peaks in the order-parameter distribution (reflecting phase coexistence),
as discussed in the context of the $J_1$-$J_2$ model in \cite{Sandvik10}.

\begin{figure}
\center{\includegraphics[width=8.4cm, clip]{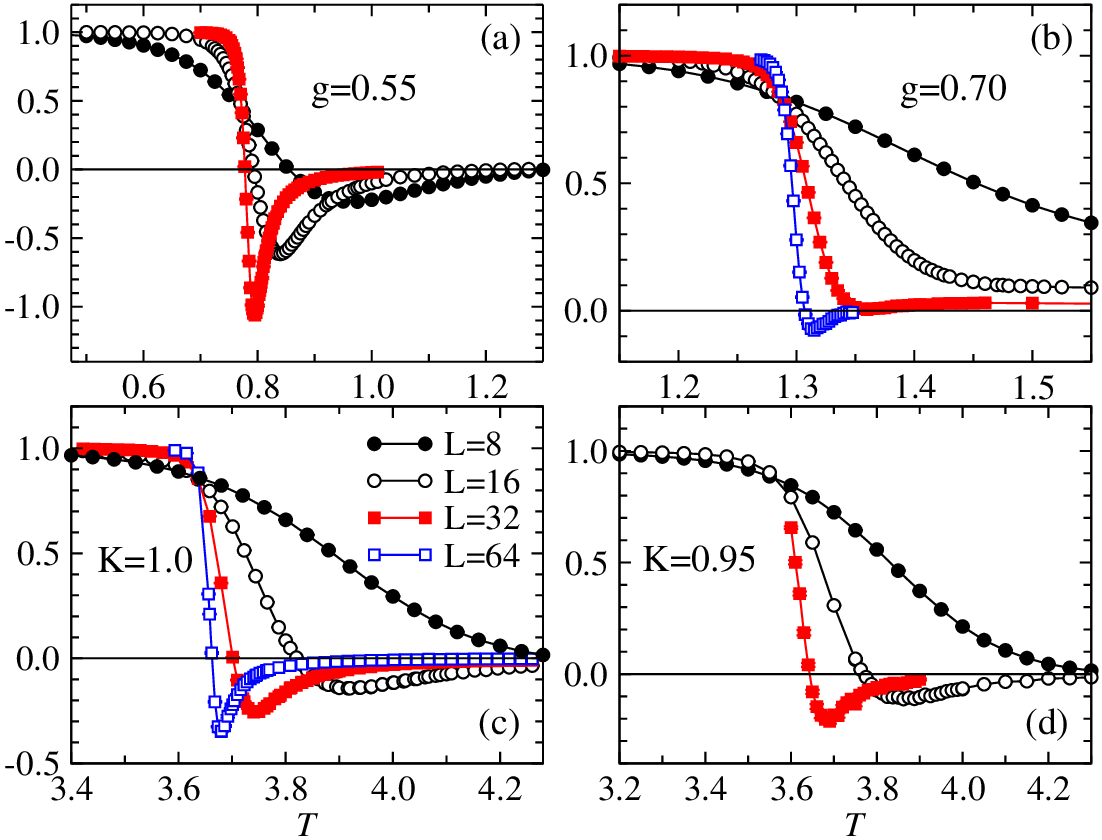}}
\vskip-1mm
\caption{(Color online) Binder cumulant vs temperature for (a,b) the $J_1$-$J_2$ model at $g=0.55$ and $0.70$, and (c,d) the AT
model at $K=1$ (the $4$-state Potts point) and $K=0.95$.}
\label{fig2}
\vskip-3mm
\end{figure}

\begin{figure}
\center{\includegraphics[width=6.8cm, clip]{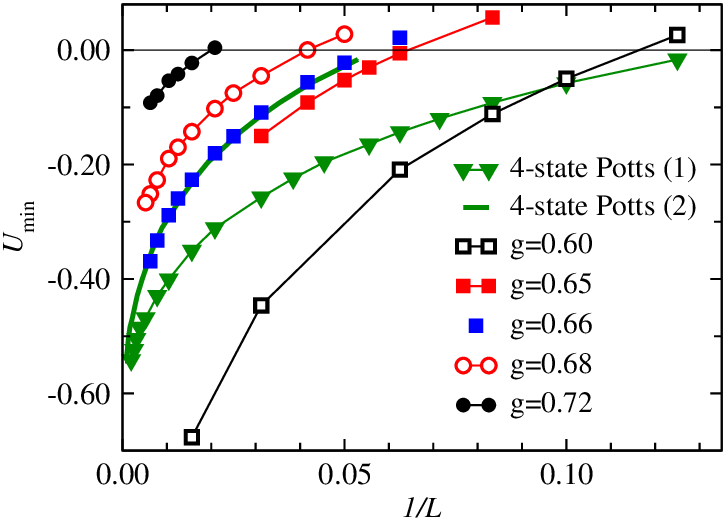}}
\vskip-1mm
\caption{(Color online) Peak value of the Binder cumulant vs $L$ for the $J_1$-$J_2$ model at different $g$ values and the $4$-state Potts model. 
For the latter (1) in the legend refers to the original data and (2) has rescaled size, $L \to aL$, with $a\approx 2.36$.}
\label{fig3}
\vskip-3mm
\end{figure}

As is clear from Figs.~\ref{fig2}(a,b), $U(T)$ of the $J_1$-$J_2$ model indeed develops a negative peak that grows with increasing $L$ for $g=0.55$ 
and $g=0.70$. As $g$ increases, the system sizes needed to observe a peak also increase, indicating a weakening discontinuity of $m$. The dependence of 
the peak value $U_{\rm min}$ on $L$ is shown in Fig.~\ref{fig3} for several values of $g$ (where $U_{\rm min}>0$ corresponds to a local minimum). 
Interpolating these data, we can extract a length $L_0(g)$ where $U_{\rm min}$ crosses zero for given $g$. $L_0(g)$ grows with increasing $g$ and 
diverges when $g \approx 0.82$ (i.e., for larger $g$ there is no negative peak). The vanishing of the negative peak might be taken as an estimate of the 
location $g^*$ of the multicritical point. One could also examine the $L(g)$ at which a minimum in $U(T)$ first forms (but is not yet negative). 
This gives a still higher value of $g^*$. However, such procedures, or ones based on multiple peaks in the order-parameter or energy distribution, 
over-estimate $g^*$ because these features can appear also for continuous transitions. Indeed, as shown in Figs~\ref{fig2}(c,d) and \ref{fig3}, we observe 
negative cumulant peaks also in the Potts and AT models, in spite of these models being rigorously known to have continuous transitions.

The most natural scenario suggested by these data is again that the $J_1$-$J_2$ model at a point $g=g^*$ corresponds to the $4$-state 
Potts model. It therefore exhibits pseudo first-order behavior for some range of $g$-values above $g^*$, just like the AT model does for $K$ at
and slightly below the Potts point $K=1$. The very similar behavior of $C_{\rm max}(L)$ for $g = 0.68$ and the $4$-state Potts model in Fig.~\ref{fig1}(a) already 
suggests $g^* \approx 0.68$. Below we will present further evidence of $g^*=0.67 \pm 0.01$ indeed being a $4$-state Potts-universal point. 

We presume that the stripe transition is first-order for $g<g^*$, instead of some alternative (and unlikely) exotic behavior outside known scenarios for Z$_4$ 
symmetry-breaking. The discontinuities are always very weak, however, so that the asymptotic first-order scaling behavior cannot be observed in practice. For 
instance, the negative $U(T)$ peak for $g<g^*$ and the specific heat appear to diverge much slower than the expected $L^2$ form. Similar behavior can also be 
observed for the $5$-state Potts model, which is a well-known prototypical example of a weak first-order transition \cite{Baxter} with anomalously 
large correlation length $\xi$ \cite{Peczak}. To observe clear first-order scaling, one has to use lattices with $L \gg \xi$.

{\it Histograms}---We next examine in some more detail the pseudo first-order signals in the AT and $J_1$-$J_2$ models, using the 
probability distribution of the order parameter and the energy. Some aspects of the full distribution $P(m_x,m_y)$ were discussed in \cite{Sandvik10}.
Here we consider $P(m^2)$. It is well known that phase coexistence at a first-order transition leads to a double-peak distribution in a narrow window
(of size $\sim 1/L^2$ in two dimensions) around $T_c$. For $L \to \infty$, the distribution approaches two delta-functions (at $m=0$ and the value 
$\bar m$ of the order-parameter just below $T_c$), with  weight shifting between the two across the narrow $T_c$ window. A double peak in the energy 
distribution corresponds to latent heat.

\begin{figure}
\center{\includegraphics[width=8cm, clip]{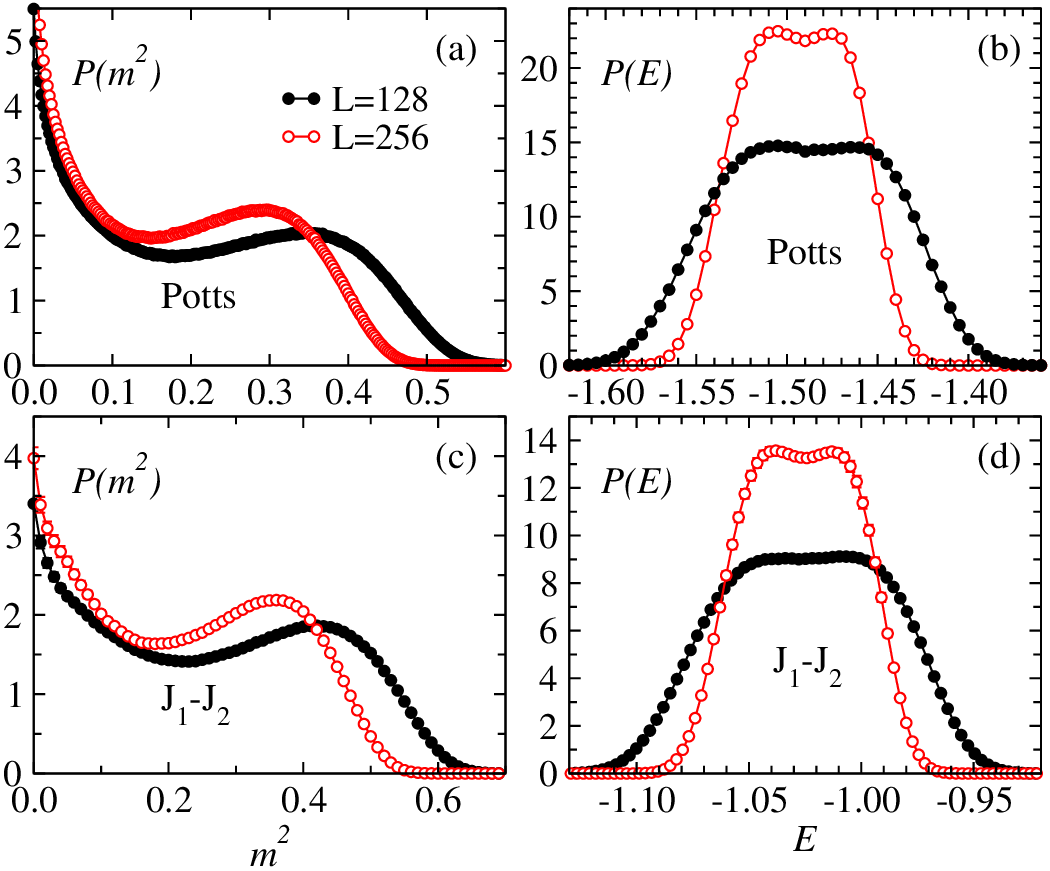}}
\vskip-1mm
\caption{(Color online) Histograms of the order parameter $m^2$ and the energy $E$ for (a),(b) the 
4-state Potts model (the $K=1$ AT model) and (c),(d) for the $J_1$-$J_2$ model 
at $g=0.67$. Here $T$ is very close to $T_c$, chosen such that the two peaks in 
the energy histograms are of the same height; for the Potts model $T/K=3.64231$ 
for $L=128$ and $3.64460$ for $L=256$, while for the $J_1$-$J_2$ model $T/J_1=1.2014$ 
for $L=128$ and $1.2004$ for $L=256$.}
\label{fig4}
\vskip-3mm
\end{figure}

As would be expected based on the negative Binder cumulants, the pseudo first-order behavior also gives rise to double peaks in the order-parameter and energy 
distributions close to the transition. This is shown in Figs.~\ref{fig4}(a,b) for the $4$-state Potts model. Similar behavior can be seen in the $J_1$-$J_2$ model, 
for which results are shown in Figs.~\ref{fig4}(c,d) at $g=0.67 \approx g^*$ (where the transition should be weakly first-order). In Ref.~\cite{honecker2} 
a double-peaked energy histogram was seen all the way up to $g=0.9$ on large lattices. This was taken as a first-order transition, while our results show that 
this is merely pseudo first-order behavior \cite{Schreiber}. Combining the results, we conclude that there is pseudo first-order behavior in the $J_1$-$J_2$ 
model from the Potts point $g^* \approx 0.67$ up to at least $g = 0.9$. 

\begin{figure}
\center{\includegraphics[width=8.2cm, clip]{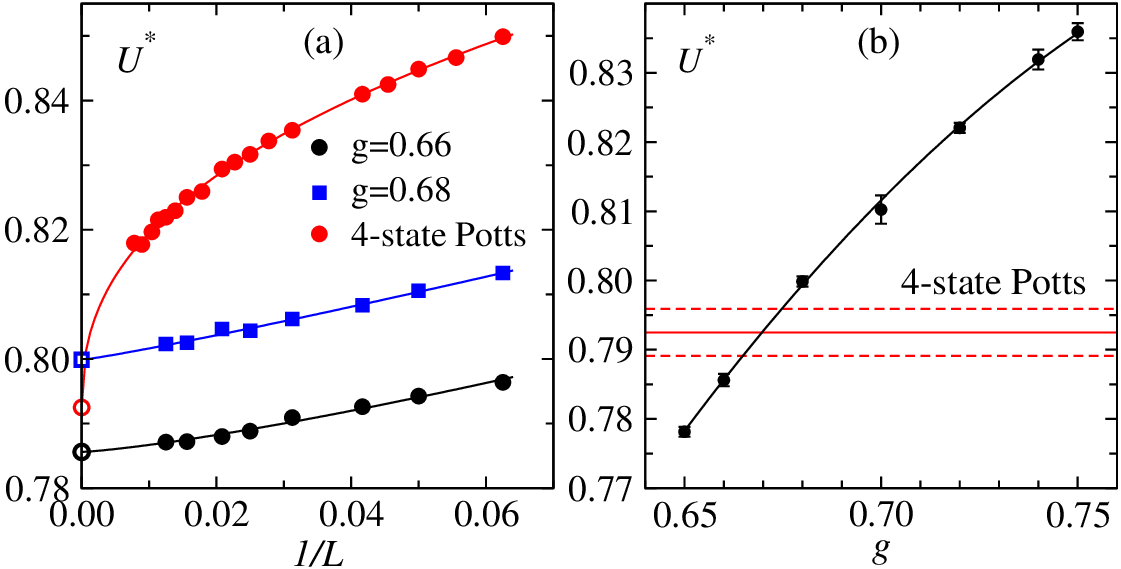}}
\vskip-1mm
\caption{(Color online ) (a) Binder cumulant crossing points for $(L,2L)$ system pairs. The approach to $U^*(L=\infty)$ is expected to be 
governed by a nonuniversal scaling correction. The curves show fits of the form $U=a+b/L^c$. (b) $L \to \infty$ extrapolated $U^*$ 
of the $J_1$-$J_2$ model compared with the Potts result (the lines indicate the estimate with error bar).} 
\label{fig5}
\vskip-3mm
\end{figure}

{\it Potts point.}---If the continuous transitions indeed belong to the AT class for $g \in [g^*,\infty]$, then a way to estimate the Potts point $g^*$ more 
precisely is to use the universal Binder crossing value $U^*$ (see Fig.~\ref{fig2}) of curves $U(T)$ for different $L$ at fixed $g$. For the $4$-state Potts 
model (the $K=1$ AT model), we estimate $U^*=0.792(4)$ by extracting crossing points between data for pairs $(L,2L)$ and extrapolating to $L=\infty$. Examples 
of the finite-size scaling of crossing points are shown in Fig~\ref{fig5}(a). In the $J_1$-$J_2$ model, $U^*(g)$ increases monotonically with $g$, as shown 
in Fig~\ref{fig5}(b). We can now estimate $g^*$ by equating $U^*(g)$ to the $4$-state Potts value. This gives $g^*=0.67 \pm 0.01$, in good agreement with the 
Potts-like behavior of the specific heat shown in Fig.~\ref{fig1}(a) and the histograms in Fig.~\ref{fig4}.

As a further test, we use a method inspired by the {\it flowgram technique} \cite{kuklov}: Since the order-parameter distribution should be
universal (up to scale factors) at $T_c$ for a continuous transition, and the negative peak in $U(T)$ represents one aspect of this distribution ($U^*$ 
being another one) the value $U_{\rm min}(L)$ should either approach a universal value (as $U^*$ does) or diverge in a universal manner when $L \rightarrow \infty$. 
Then, $U_{\rm min}(L)$ for the $4$-state Potts model and the $J_1$-$J_2$ model at $g^*$ should, if the models are controlled by the same fixed 
point, collapse onto the same curve for large systems, once a rescaling $L \rightarrow aL$ is introduced for one of the models. This is indeed the case if, 
and only if, $g$ is close to the $g^*$ point estimated above. Scaled Potts data match best the $g=0.66$ curve in Fig.~\ref{fig3}. It is not clear whether 
$U_{\rm min}$ here diverges very slowly or converges to a finite value.

{\it Conclusions.}---By combining many mutually consistent signals, we have demonstrated a point $g=g^*=0.67 \pm 0.01$ at which the $J_1$-$J_2$ model is 
controlled by the $4$-state Potts fixed point. The scenario of the critical curve for $g \in [g^*,\infty)$ being in one-to-one correspondence with that of
the AT model for $K \in [1,0)$ is the only possibility consistent with known CFT scenarios and MC results. The pseudo first-order behavior uncovered
here, along with the very weak first-order transitions for $g \in (1/2,g^*)$, is what made it so difficult to correctly  characterize 
the nature of the transitions until now. 

With the nature of the transition now understood, it will be interesting to study other aspect of the model, i.e., the kinetics \cite{Shore,Dominguez}
at the weakly first-order and pseudo first-order transitions. We also note that many interesting quantum problems involve the same kind of order-parameter 
symmetry in two dimensions, e.g., stripe states in models of high-$T_c$ superconductors \cite{White,DelMaestro} and valence-bond-solid states in quantum 
antiferromagnets \cite{Lou}. The $T>0$ transitions in these systems, as well, may be impacted by the issues we have pointed out here.
{\it Acknowledgments.}---We thank Bill Klein and Sid Redner for useful discussions. 
This work was supported in part by the NSF under Grant No.~DMR-1104708. 

\null

\end{document}